\documentstyle[twocolumn,aps,floats]{revtex}

\begin{document}

\draft \tolerance = 10000

\setcounter{topnumber}{1}
\renewcommand{\topfraction}{0.9}
\renewcommand{\textfraction}{0.1}
\renewcommand{\floatpagefraction}{0.9}
\newcommand{\br}{{\bf r}}

\twocolumn[\hsize\textwidth\columnwidth\hsize\csname
@twocolumnfalse\endcsname

\title{Why We Can Not Walk To and Fro in Time as Do it in Space?
 (Why the Arrow of Time is Exists?)}
\author{L.Ya.Kobelev\\
 Department of  Physics, Urals State University \\ Lenina Ave., 51,
Ekaterinburg 620083, Russia  \\ E-mail: leonid.kobelev@usu.ru}
\maketitle

\begin{abstract}
Existence of arrow of time  in our world may be easy explained if time has
multifractal nature. The interpretation of nature of time arrow is made on
the base of   multifractal theory of time and space presented at works
\cite{kob1}-\cite{kob15}. In this paper shown possibility  to walk to and
fro in space and necessity of huge amount of energy for stopping time and
changing direction of it in microscopic volumes.  $$ CONTENTS: $$ 1.
Introduction\\2. Universe as Time and Space with Fractional Dimensions
\\3. Why  Time has Direction Only to Future and Why Impossible to Walk in
Time To and Fro?\\4. Is It Possible to Change Direction of Time and How
Much Energy It Needs? \\5.How Much Energy Needs for Stopping  Time and
Moving it Back in the Volume of  Cubic Centimeter During One Second?\\6.
Why We Can Walk To and Fro in Our Space?\\7. Conclusions
\end{abstract}

 \vspace{1cm}

]
\section {Introduction}

As is well known  the time flows in our world only in future (arrow of
time of Eddington \cite{ed} and Prigogin\cite{prig1}) and now nobody knows
is it possible to change direction of the time flow or is it impossible
and why if it is possible or impossible. The problems of arrow of time are
very intrigues   and many of physicians presented interesting models of it
phenomenon (see for example \cite{granic}, or new interesting experiments
\cite{mug} . For analyzing this problem it is necessary to investigate the
models of time in which the time is not simple time axes, but  has
complicated structure. The purpose of this paper is to investigate problem
of time arrow in the multifractal model of time presented in fractal model
of space and time at \cite{kob1}-\cite{kob15}. In this model the time and
the space are real material fields  with fractional dimensions and
multifractal structure (multyfractal sets) defined on sets of their
carriers of measure. In every time (or space) points the dimensions of
time (or space) determine densities of Lagrangians energy for all known
physical fields ( or Lagrangian of new physical fields for space
dimensions) in these points. This model allows understand reason of
existence of the arrow of time: this reason has pure energetically nature
and consists in necessity to use huge amount of energy for changing of
direction of time.

\section{Universe as  time and space with fractional dimensions}

In this paragraph we summary the main results of the fractal theory of
time and space \cite{kob1}-\cite{kob15}. In this theory,  when Universe
was born ( the moment of "big bang") from vacuum (in this theory Universe
was born from the set of carrier of measure which plays role of physical
vacuum) only material fields were born (or appear on surface of carrier of
measure-vacuum): the time and the space fields with fractional dimensions.
So our Universe are real fields of time and space and not conclude
something yet. Fractional dimensions of time and space are appearing in
our world as  physical fields (all physical fields, known and new fields
that will be find, are characteristics of  time with fractional
dimensions). When temperature of Universe (i.e. the temperature of fields
of time and space) was changing, fractional dimensions was changing too
and new physical fields were appearing in accordance with known theory of
"big bang" and broken symmetry ( depending at Universe temperature). The
equations for physical fields appear as consequences of principle of
fractal dimensions functional (FDF) minimum . These equations are Euler
equations with generalized fractional Riemann-Liouville derivatives (the
generalization consists in propagating the Riemann-Liouville derivatives
on domain of multifractal sets where fractional dimensions are functions
of time and coordinates) . For case of integer dimensions these
derivatives and equations coincide with ordinary derivatives and the
theory in whole coincides with known physical theories. So only difference
of the theory \cite{kob1}-\cite{kob15} from known theories consists in
using the algorithm for propagating modern theories on the domain of the
Universe with fractional dimensions of time and space and geometrization
of all physical fields in frame of fractal geometry of world. In cited
works many equations of modern physics were researched in fractal space
and time. It was shown that in multifractal model all physical fields are
geometrizationed. It differs this theory positively  from general
relativity theory (the latter is a special case of multifractal theory of
time and space in special selection of measure carrier and integer
dimensions (see \cite{kob9}) ) where only one field (gravitation field) is
geometrizationed.\\ In the world of fractional dimensions there are many
new special physical characteristics and peculiarities. The main of them
are:\\ a) in the world with fractal dimensions there are no constant
physical values because fractional derivative with respect to constant
value is not zero. So all physical values are changing in time and in
space ( for example, electric charge of electron has changed from the time
of big bang as $\delta e=e\varepsilon/t $ where $t=t_{0}+\delta t$ and
$\delta t$ is current time, $t_{0}$ -is the time of existence of
Universe). Consequences of it are absence in this world any rigorous laws
of preservation  (they are fulfilled only as very good approach, but not
as rigorously laws);\\ b) there are no inertial systems in Universe
because there are no constant velocities and special Einstein theory must
be replaced by theory of almost inertial systems (it coincide with special
relativity theory on the surface of Earth till velocities $v=c- \delta v,$
$ \delta v\sim 100msec^{-1}$; On the surface of Earth the fractional
dimensions of time differs from unity on value $\sim 10^{-9}$.
\\c) all systems of reference are absolute systems (time and space are  non
homogeneous and non isotropic). So Michelson experiments had proved
independence of speed of light at moving of origin of light are only very
good approach ( on the surface of Earth the changing of direction $v$ on
$-v$ in the fractal world gives change of speed of light on value $\sim
0.1 cmsec^{-1}$ ); \\d) in the fractal world velocities of moving bodies
may be equal speed of light in vacuum (there are no singularity at $v=c$)
or exceed it and it is possible to  propagate information with any wished
velocities; \\ e) there are no singularities in the theory because for
$V(r)\sim \infty$ ($V$ is a potential of any fields) all fractional
differential operators of the theory turns into fractional integrals;\\f)
all physical (i.e. Newton ,Shr\"{o}dinger, Maxwell, Dirac, Einstein and so
on ) equations of modern physics are  irreversible;\\g) the laws of
thermodynamics are consequences of multifractal structure of Universe for
its domain with state near thermodynamics equilibrium; \\h) the theory
predicts existence of new fields originated by fractional dimensions of
space ;\\i) the theory predicts existence of new physics in the domain of
superluminal velocities in vacuum for ordinary (not taxions) particles
with new effects that may be experimentally discovered;\\j) in the fractal
world the time has inertial characteristics described by $m_{r}$ in
analogy with well known Newton mass $m_{t}$ and equations for $t({\mathbf
r})$ are exist (analogies of Newton, Dirac and so on equations) ;\\ k) the
theory have two masses: masses $m_{t}$ as measure of inertia for moving in
the time(Newton masses) and masses $m_{r}$ as measure of inertia for
moving time in space (masses concerned with inhomogeneous time in the
space);
\\l) the presented theory is a natural generalization of all modern physical
theories for domains of time and space with fractional dimensions and
coincide with any of them in the case when fractional parts of dimensions
are zero. This theory is not the  variant of theories of quantum time and
quantum space, because the multifractal "intervals" of space and time had
used in the theory (the time and the space are consist of them) are very
composed multifractal sets and researching its structure lay in f
uture.\\The question about irriversibility of time (the arrow of time) in
the theory of fractal time and space (\cite{kob1}-\cite{kob12}) was not
researched. This paper has purpose to investigate why the time has only
one direction in our world on the base of the  multifractal theory of time
and space \cite{kob1}-\cite{kob15}.

\section{Why  time has direction  only to future and why impossible to walk
in time to and fro?}

In the theory of fractal time and space the problem of existing in
Universe the arrow of time may be considered (we show it below) as the
problem of decreasing of the energy for the states when  the time arrow
has direction to future. We will show that in the domain of multifractal
Universe with time dimensions less than unit and when the fractional parts
of time dimensions are small additions (with negative signs) to unit , the
arrow of time with direction to future gives decreasing of energy for any
body in this domain of Universe. So the arrow of time gives spontaneous
decreasing (diminishing) of Universe energy in these domains (and in
Universe on the whole).\\ For demonstrating it let us write the quantum
equations for model particle with a rest mass $m$ and momentum ${\mathbf
p} =0$ for two cases: in the time space with integer dimensions and in the
time space with fractional dimensions
\begin{equation}\label{1}
  i\hbar\frac{\partial}{\partial t}\psi({\mathbf r},t) - mc^{2}\psi({\mathbf r},t) = 0
\end{equation}
\begin{equation}\label{2}
 i\hbar D_{0,+}^{d_{t}} \psi({\mathbf r},t) - m^{2}\psi({\mathbf r},t) = 0
\end{equation}
In (\ref{2}) we used generalized fractional derivative $D_{+,t}^{d_{t}}$
defined as (see \cite{kob1}-\cite{kob15}). Following these works we
consider both time and space as the initial real material fields existing
in the world and generating all other physical fields by means of their
fractal dimensions. Assume that every of them consists of a continuous,
but not differentiable bounded set of small intervals (these intervals
further treated as "points"). Consider the set of small time intervals
$S_{t}$ (their sizes may be evaluated in rude approach as Planck sizes).
Let time be defined on multifractal subsets of such intervals, defined on
certain measure carrier $\mathcal{R}^{N}$. Each interval of these subsets
(or "points") is characterized by the fractional (fractal) dimension (FD)
$d_{t}({\mathbf r}(t),t)$ and for different intervals FD are different. In
this case the classical mathematical calculus or fractional (say, Riemann
- Liouville) calculus \cite{sam} can not be applied to describe a small
changes of a continuous function of physical values $f(t)$, defined on
time subsets $S_{t}$, because the fractional exponent depends on the
coordinates and time. Therefore, we have to introduce integral functionals
(both left-sided and right-sided) which are suitable to describe the
dynamics of functions defined on multifractal sets (see
\cite{kob1}-\cite{kob3}). Actually, these functionals are simple and
natural generalization of the Riemann-Liouville fractional derivatives and
integrals:
\begin{equation} \label{3}
D_{+,t}^{d}f(t)=\left( \frac{d}{dt}\right)^{n}\int_{a}^{t}
\frac{f(t^{\prime})dt^{\prime}}{\Gamma
(n-d(t^{\prime}))(t-t^{\prime})^{d(t^{\prime})-n+1}}
\end{equation}
\begin{equation} \label{4}
D_{-,t}^{d}f(t)=(-1)^{n}\left( \frac{d}{dt}\right)
^{n}\int_{t}^{b}\frac{f(t^{\prime})dt^{\prime}}{\Gamma
(n-d(t^{\prime}))(t^{\prime}-t)^{d(t^{\prime})-n+1}}
\end{equation}
where $\Gamma(x)$ is Euler's gamma function, and $a$ and $b$ are some
constants from $[0,\infty)$. In these definitions, as usually, $n=\{d\}+1$
, where $\{d\}$ is the integer part of $d$ if $d\geq 0$ (i.e. $n-1\le
d<n$) and $n=0$ for $d<0$. If $d=const$, the generalized fractional
derivatives (GFD) (\ref{1})-(\ref{2}) coincide with the Riemann -
Liouville fractional derivatives ($d\geq 0$) or fractional integrals
($d<0$). When $d=n+\varepsilon (t),\, \varepsilon (t)\rightarrow 0$, GFD
can be represented by means of integer derivatives and integrals. For
$n=1$, that is, $d=1+\varepsilon$, $\left| \varepsilon \right| <<1$ it is
possible to obtain:
\begin{eqnarray}\label{5}
D_{+,t}^{1+\varepsilon }f({\mathbf r}(t),t)\approx
\frac{\partial}{\partial t} f({\mathbf r}(t),t)+ \nonumber \\ +
a\frac{\partial}{\partial t}\left[\varepsilon (r(t),t)f({\mathbf
r}(t),t)\right]+ \frac{\varepsilon ({\mathbf r}(t),t) f({\mathbf
r}(t),t)}{t}
\end{eqnarray}
where $a$ is a $constant$ and determined by  choice of the rules of
regularization of integrals (\cite{kob1})-(\cite{kob2}), (\cite{kob7})
(for more detailed see \cite{kob7}) and the last addendum in the right
hand side of (\ref{5}) is very small. The selection of the rule of
regularization that gives a real additives for usual derivative in
(\ref{3}) yield $a=0.5$ for $d<1$ \cite{kob1}. The functions under
integral sign in (\ref{3})-(\ref{4}) we consider as the generalized
functions defined on the set of the finite functions \cite{gel}. The
notions of GFD, similar to (\ref{3})-(\ref{4}), can also be defined and
for the space variables ${\mathbf r}$. The definitions of GFD
(\ref{3})-(\ref{4}) needs in connections between fractal dimensions of
time $d_{t}({\mathbf r}(t),t)$ and characteristics of physical fields
(say, potentials $\Phi _{i}({\mathbf r}(t),t),\,i=1,2,..)$ or densities of
Lagrangians $L_{i}$) and it was defined in cited works. Following
\cite{kob1}-\cite{kob15}, we define this connection by the relation
\begin{equation} \label{6}
d_{t}({\mathbf r}(t),t)=1+\sum_{i}\beta_{i}L_{i}(\Phi_{i} ({\mathbf
r}(t),t))
\end{equation}
where $L_{i}$ are densities of energy of physical fields, $\beta_{i}$ are
dimensional constants with physical dimension of $[L_{i}]^{-1}$ (it is
worth to choose $\beta _{i}^{\prime}$ in the form $\beta _{i}^{\prime
}=a^{-1}\beta _{i}$ for the sake of independence from regularization
constant). The definition of time as the system of subsets and definition
of the FD for $d^{t}$ (see ( \ref{4})) connects the value of fractional
(fractal) dimension $d_{t}(r(t),t)$  with each time instant $t$. The
latter depends both on time $t$ and coordinates ${\mathbf r}$. If
$d_{t}=1$ (an absence of physical fields) the set of time has topological
dimension equal to unity. The multifractal model of time allows ( as was
 shown \cite{kob5}) to consider the divergence of energy of masses
moving with speed of light in the SR theory as the result of the
requirement of rigorous validity of the laws pointed out in the beginning
of this paper in the presence of physical fields (in the multifractal
 theory there are only approximate fulfillment of these laws). We bound
consideration only the case when relation $d_{t}=1-\varepsilon({\mathbf
r}(t),t))$, $|\varepsilon|\ll 1$ are fulfilled.  In that case the GFD may
be represented (as a good approach) by ordinary derivatives and relation
(\ref{5}) are valid. So the equation (\ref{2}) reeds
 \begin{eqnarray}\label{7}
 i\hbar \frac{\partial}{\partial t} \psi({\mathbf r},t) - mc^{2}\psi({\mathbf r}(t),t)+\nonumber \\
 + i\hbar\frac{\partial}{\partial t}[\varepsilon \psi({\mathbf r}(t),t)]+
i\hbar \frac{\varepsilon \psi}{t} = 0
\end{eqnarray}
This equation describes behavior of the particle with point sizes in time
and space (we  remind that it is only  the approach that we use and in
reality minimal size of time intervals and minimal sizes of  space
intervals in the theory  are bound, for example, by Planck sizes, thou the
last are multifractal sets too) For free ( more rightly almost free)
particle choose solution for $\psi$ as a plane wave with energy depending
of time ($\psi=\psi_{0}\exp\frac{-iE(t)}{\hbar} $) and for domain of
time-space where by members with $\frac{\partial \varepsilon}{\partial t}$
may neglect ( i.e. fractional additives almost constant) receive
\begin{equation}\label{8}
\psi(t)= \psi_{0}\exp(\frac{-i}{\hbar}E(t) t)
\end{equation}
\begin{equation}\label{9}
    E(t) = mc^{2} + \pi\hbar \frac{\varepsilon }{t}-i\frac{\hbar\varepsilon\ln t}{t}
\end{equation}
or
\begin{equation}\label{10}
 \psi(t)= \frac{\psi_{0}}{t^{\varepsilon}}\exp(\frac{-i}{\hbar}\tilde{E}(t) t)
\end{equation}
where
\begin{equation}\label{11}
  \tilde{E}= mc^{2} + \frac{\varepsilon \hbar}{t}
\end{equation}
The equations (\ref{10}) - (\ref{11}) allow to conclude that the fractal
dimensions time leads to the two sorts of phenomena: a) decreasing of
energy with time flow on value $\varepsilon\hbar t^{-1}$; b) spread (run)
of wave function on value $t^{-\varepsilon}\sim (1-\varepsilon \ln t) $.
Thus when time $t$ (or current time $(t-t_{0})$) is increasing ($t_{0}$ is
age of Universe) both energy and wave function are decreasing. Evaluation
of both decreasing values (if take into account only gravitation field on
surface of Earth and take into account that $\varepsilon\sim 10^{-7}$,
$t\sim t_{0}\sim 10^{17}$) gives: $\triangle E=\hbar\varepsilon
t_{0}^{-2}\sim10^{-56}ev$ for one second of current time . We pay
attention that so little value of damping energy follows from the points
model for describing the particle (the particle consists are of one
"interval" of time and of one "interval" of space, described above). For
more realistic model it is necessary to take into account a really sizes
of particles ( partly the energy of damping will be evaluates in next
paragraph). Rigorously say we made evaluation of the lose energy in the
point of space where point  particle presents because  the fractional
dimensions that are sources of the real particle mass may be so large that
approach  $\varepsilon <<1$ is not work.

\section{Is it possible to change the direction of time and how much energy it
 needs ?}

In this paragraph we research the question: may the time be turned in back
direction? The energies needs for is possible to evaluate if use rude
approach . The example for behavior of the free model particle has
demonstrated the damping of energy in Universe with fractional time
dimensions ( the case of decreasing energy as consequences of existence of
fractional dimensions of space is analogies). Thus the question why the
time has only one direction towards future has natural answer: it is
because only in that direction of time the energy of particles ( or any
bodies consisting of particles)  decrease. If somebody wants to change the
direction of time it is necessary to spend energy for changing structure
of real fields of time and space. On the first look this energy is very
small (see above paragraph), but its smallness is related with the case
when multifractal "intervals" of time and space ($\triangle t$ and $
\triangle {\mathbf r} $) were treated as "points" with fractional (global
for sets consisting the time field) dimensions and equation (\ref{11})
describes the lose energy only  of such points "intervals". What are
values of time and space "intervals " in our Universe? The theory of
fractal time and space in her present state can not answer on this
question. So we  use some hypotheses about their values. As the rude
approach we may take for its values Planck sizes:$ \triangle t \sim
10^{-44}sec$  and  $ \triangle x\sim 10^{-33}cm $ . Then one second
consists of $10^{44}$ of "intervals" of time and one centimeter consists
of $10^{33}$ "intervals" of space ( we needs to remember that every of
"intervals" is multifractal sets with very composed characteristics which
do not researched in present work (see \cite{kob5}). For current time $
(t-t_{0})$ the relation  for lose energy by the domain space with volume
$\sim10^{-42}cm^{3} $ during one second ( the volume of elementary
particles) reads
\begin{equation}\label{12}
 \triangle E \sim \frac{\varepsilon\hbar}{t_{0}^{2}}10^{44}10^{-42}10^{99} \sim
 10^{45}ev
\end{equation}
If  such gigantic energy  will be  received by time field with space
volume $10^{-42}cm^{3}sec$ the flow of time during one second be stopped
and if double this energy during one second the time flow change its
direction (i.e. $t\rightarrow -t$) and time will flow one second in back
direction). Thus the direction of time in our Universe may be changed but
it needs in the gigantic amount of energy. Of course, the value of this
energy depends at the evaluation of the "intervals" of time and space
values and if last values more than Planck intervals (for example at
$10^{5}- 10^{10}$) the energy will be smaller but also huge. Result is: in
principle the inversion of direction of time may be reached but it is
impossible on the modern state of humankind technology even for
microscopic volumes. If values of time and space intervals needs in
corrections, the evaluation of energy needs for inversion of time
direction must be corrected too.

\section{How much energy needs for stopping the time and moving it back in
the volume of one cubic centimeter during one second?}

In the multifractal fractal theory of time and space where time and space
fields are real themselves and are real origin of all physical fields in
principle (it was shown in above paragraph) there are possibilities to
inverse the time flow in back direction. Now we write the equation for
changing with time of one Planck interval of space $x_{p}$ and see how the
energy of it (as a part of real space field it has energy of rest which
damping with flow of time) changes in time (let this Planck volume is in
rest, i.e. ${\bf p}=0)$
\begin{equation}\label{13}
  i\hbar D_{+,t}^{d_{t}}x_{p}=E_{0}x_{p}
\end{equation}
or
\begin{equation}\label{14}
 i\hbar \frac{\partial}{\partial t}x_{p}= E_{0}x_{p} - i\hbar
 \frac{\varepsilon}{t}x_{p}
\end{equation}
 The solution of (\ref{14}) may be  represented as
\begin{equation}\label{15}
 x_{p}=x_{0}\exp{-\frac{i}{\hbar}E(t)t}
\end{equation}
where
\begin{equation}\label{16}
  E(t)=E_{0}+ \frac{\varepsilon \hbar}{t}
\end{equation}
Now evaluate the volume $x_{p}$ using Planck interval and use earlier
values of $\varepsilon, t_{0}, \hbar$.  If connection binding every
element ${\bf r}$ with element $t$ has the form $d{\bf r}^{2} - dt^{2} =
0$ and for each element of space spending energy each element of time
spend energy too, thus for the energy lose of space volume equal one
$cm^{3}$ during one second write
\begin{equation}\label{17}
\triangle E\sim \frac{\varepsilon}{t_{0}^{2}}10^{99}10^{44}
  10^{12}ev\sim 10^{-68}10^{155}ev\sim10^{87}ev
\end{equation}
So we got the order of values of energy needs for stopping the time in the
$cm^{3}sec$. For inversion of time flow in this volume needs double this
value of energy. For stopping  time in the volume of one elementary
particle necessary multiply above value at $10^{-42}$ (if size of particle
$\sim 10^{-14}cm$). It gives $ \triangle E\sim10^{45}ev$. Nobody knows is
it  possible in far future to receive such energies and concentrate them
in small volumes.

\section{Why we can walk to and fro in our space}

Why we can not walk to and fro in time had been explained in the frame of
multifractal time in paragraph above on the language of energetically
reasons. The possibility of walking to and fro in space is conditioned by
vector characteristics of fractional addendum to space derivatives in
multifractal Universe. For simplicity we consider non relativistic case
when particle is described by Shr\"{o}dinger equation in multifractal
space (see \cite{kob1}, \cite{kob2}, \cite{kob12}) .  Let multifractional
addendum to integer space dimensions $\varepsilon_{i}$ is very small
($|\varepsilon_{i}|<<1$). Than for GFD we can right (we conserve only main
addendum  necessary for our purpose)
\begin{equation}\label{18}
D_{+,{\mathbf r}}^{1-\varepsilon_{{\mathbf r}}} \sim
(\frac{\partial}{\partial {\mathbf r}}+ \frac{\varepsilon_{\mathbf
r}\hbar}{{\mathbf r}})
\end{equation}
and for Shr\"{o}dinger equation in fractal space for free particle receive
\begin{equation}\label{19}
 i\hbar D_{+,t}^{1-\varepsilon_{t}}\psi=
  - \frac{\hbar^{2}}{2m}D_{-,{\mathbf r}}^{1-\varepsilon_{\mathbf r}}
 D_{+,{\mathbf r}}^{1-\varepsilon_{{\mathbf r}}}\psi
\end{equation}
Now replace fractional space derivatives by means of (\ref{13}) then
(\ref{14}) reads (if neglect the members of order $\varepsilon^{2}$ and
non essential scalar members in right hand part of equation )
\begin{equation}\label{20}
i\hbar D_{+,t}^{1-\varepsilon_{t}}\psi=  -
\frac{\hbar^{2}}{2m}\triangle\psi +\nabla(\frac{\varepsilon_{{\bf
r}}\hbar{\mathbf }}{{\bf r}})\psi
\end{equation}
If in the (\ref{20}) we  replace ${\bf r}$ by $-{\bf r}$ the sign of
fractal addendum from fractional space dimensions do not change its sign,
so there are no energetically reasons forbidding walking to and fro in the
fractal Universe. Of course, in the equation (\ref{15}) omitted the
members describing the lose of energy by particle reasoned by the
fractional structure of space. We do not evaluate the energy lose reasoned
by multifractal structure of space  because  in this case the evaluation
is very difficult ( the value of $\varepsilon_{{\bf r}}$ is unknown). It
value defined by  new fields (not discovered yet) that borne by fractional
space dimensions (see \cite{kob1}, \cite{kob2}, \cite{kob3}.

\section{conclusions}

The arrow of time in considered model of multifractal time and space as
was seen above is consequence of energetically reasons. Direction of time
may be changed (thou only in principle in our epoch because of huge amount
of energy that needs for it). There are three main results of this
article: a) the explanation of nature of arrow of time by natural lose of
energy of our Universe and by necessity for  energy  compensation of this
lose for changing of direction of time in any domain (small or large) of
space and time; b) it is point out at the huge amount of energy in every
bodies and fields ( more detailed consideration will be in special paper )
caused by the real nature of fields of time and space ; c) the principle
possibility to change directions of  time and space fields in the remote
future epoch. We considered the energy needs for changing direction of
time, thou it is necessary to return the space in earlier state too.\\
Some general remarks . Any fractal or multifractal sets (Universe is
multifractal set) always not belong and not coincide with measure of
carrier on that they are  defined (it include the cases when measure of
carrier is multifractal set itself and not space $R^{N}$ type ). If
describe the measure carrier of our Universe in terms of "physical
vacuum", then "vacuum" do not belongs to our Universe. Main part of it
lays out of Universe (see also \cite{granic}) . The Universe may be
treated as a gigantic energetic fluctuation ("metastable" long living
fluctuation) in the measure of carrier and as the fluctuation it has
strong binding with its "mother" . The existence of strong binding with
the vacuum (measure of carrier) consist in continual transferring to
vacuum the huge amount of energy that was got from vacuum in the moment of
"big bang"  (or in the moment of birth in any over scenario) as was
demonstrated in this paper. What future wait our Universe in the model of
multifractal space and time ? Universe will spend her supply of energy
that was got from vacuum in the moment of big bang. When all energy supply
be spent process of Universe dying will be finished till time when new
Universe borne from measure of carrier (vacuum). In this model many of
universes (may be infinity) may existent because measure of carrier can
give birth any huge amount of Universes (with their own times and spaces
that can different at our time and space by it dimensions and energetical
characteristics) and die of one of them is not essential for carrier of
measure in this model of multi Universes structure of "vacuum " not
belonging to our Universe. How name this world of infinity of Universes
where birth and die of infinity of Universes change one another ? May be
"perpetual universes eternity model" will useful enough? I do not know.

\end{document}